\documentclass[12pt]{article}

\pdfoutput=1

\usepackage{cite}
\usepackage{mathrsfs,amsmath,bm,fixmath}

\catcode`@=11

\let\@CITE=\@cite
\let\Cite=\cite
\def\@cite#1#2{{#1\if@tempswa , #2\fi}}

\def\bcite{\Cite}
\def\cite#1{[\Cite{#1}]}

\catcode`@=12

\evensidemargin=\oddsidemargin
\def\Eqn#1{Eq.\ (\ref{#1})}
\def\Eqs#1#2{Eqs.\ (\ref{#1}) and (\ref{#2})}
\def\next{\nonumber\\}
\def\Sect#1{Sec.\,\ref{#1}}
\def\textcite#1{Ref.\ \cite{#1}}

\def\lag{{\mathscr L}}
\def\tilde{\widetilde}
\def\sup#1{^{\rm #1}}
\def\bsup#1{^{(#1)}}
\def\bsub#1{_{(#1)}}
\def\mat#1{\mathchoice
{\mathop{\mbox{\boldmath $#1$}}}
{{\mbox{\boldmath $#1$}}}
{{\mbox{\boldmath $\scriptstyle #1$}}}
{\mbox{\boldmath $\scriptscriptstyle #1$}}} %

\def\mat#1{\bm #1}

\title{\bf Photon propagation in non-trivial backgrounds}

\author{\bf Palash B. Pal \\
Physics Department, University of Calcutta,\\ 92 APC Road, Calcutta
700009, India}

\date{May 2020}

\begin{document}

\maketitle

\begin{abstract}
  Propagation of photons (or of any spin-1 boson) is of interest in
  different kinds of non-trivial background, including a thermal bath,
  or a background magnetic field, or both.  We give a unified
  treatment of all such cases, casting the problem as a matrix
  eigenvalue problem.  The matrix in question is not a normal matrix,
  and therefore care should be given to distinguish the right
  eigenvectors from the left eigenvectors.  The polarization vectors
  are shown to be right eigenvectors of this matrix, and the
  polarization sum formula is seen as the completeness relation of the
  eigenvectors.  We show how this method is successfully applied to
  different non-trivial backgrounds.

\end{abstract}

\bigskip

\section{Introduction}
Propagation of electromagnetic waves through any background is a
subject of huge interest.  In the language of quantum theory, we
rename this subject as the question of photon propagation.  Using the
methods of quantum field theory in thermal background, the question of
photon propagation was analyzed in a thermal background
\cite{Kislinger:1975uy, Weldon:1982aq}, and well-known attributes of
material medium like the dielectric constant and the magnetic
permeability were identified in the framework of the quantum field
theoretical treatment.  It was also shown how to extend this analysis
to chiral media \cite{Nieves:1988qz} and describe natural optical
activity in quantum field theoretic terms.  The propagation of photons
in pure magnetic fields was discussed later \cite{Ganguly:1999ts,
  Shabad:2010hx, Hattori:2012je}, and an expression for the Faraday
effect was derived in terms of form factors that appear in the quantum
theoretical framework.  The more general case of a magnetic field in a
medium, combining the two kinds of backgrounds mentioned earlier, has
also been a subject of great interest \cite{Ganguly:1999ts,
  Karmakar:2018aig}.  In many cases the discussion does not explicitly
mention photons, but rather gluons \cite{Bordag:2008wp,
  Hattori:2017xoo, Nopoush:2017zbu} or even $\rho$-mesons
\cite{Ghosh:2019fet}, but that makes no difference in the general
structure of the problem.  Basically, it is the problem of the
propagation of a spin-1 boson in a non-trivial background.

For each kind of background, one introduces appropriate parameters and
notations, and obtains the photon propagator and dispersion relations
from there.  What we propose to do in this paper is to develop a
unified approach that works for all kinds of background.  This will
give us important insight into the question of photon propagation,
which we will then apply to specific backgrounds, rediscovering some
of the old formulas with the new insight, and finding expressions for
the polarization vectors which are sometimes difficult to find from
the usual approach.

\section{The self-energy function}\label{s:se}
The momentum-space Lagrangian of a system is related to the action by
the relation
\begin{eqnarray}
  {\mathscr A} = \int{d^4k \over (2\pi)^4} \; \lag(k) \,.
  \label{action}
\end{eqnarray}
For the photon field in the vacuum, the momentum-space Lagrangian
is given by
\begin{eqnarray}
  \lag_0(k) = - \frac12 k^2 \tilde\eta_{\mu\nu} A^\mu(-k) A^\nu(k) -
  A^\mu(-k) j_\mu(k) \,, 
  \label{L0}
\end{eqnarray}
where
\begin{eqnarray}
  \tilde\eta_{\mu\nu} = \eta_{\mu\nu} - {k_\mu k_\nu \over k^2} \,.
  \label{etatil}
\end{eqnarray}
Quantum corrections in the vacuum give an extra contribution which has
the same generic form, and therefore can be absorbed in the definition
of the photon field.  In a non-trivial background, however, the photon
Lagrangian will have non-trivial contributions, and will therefore be
modified to
\begin{eqnarray}
  \lag = \lag_0 + \lag' \,,
  \label{L0+L'}
\end{eqnarray}
where the quadratic part of $\lag'$ can be written as
\begin{eqnarray}
  \lag'(k) = \frac12 \Pi_{\mu\nu}(k) A^\mu(-k) A^\nu(k) \,. 
  \label{L'}
\end{eqnarray}
Here, $\Pi_{\mu\nu}(k)$ represents the self-energy function.  In
addition to the momentum $k$, it can depend on other parameters that
characterize the background.  Those parameters depend on the nature of
the background, and are not shown here in order to maintain
generality.

In order to proceed, we need to know some properties of the
self-energy function.  We itemize them now.
\begin{description}
\item [Gauge invariance~:] Gauge invariance says that adding a term
  proportional to $k^\mu$ to the photon field should not have any
  physical implication.  This property gives the relations
  \begin{subequations}
    \label{kPi}
  \begin{eqnarray}
    k^\mu \Pi_{\mu\nu}(k) &=& 0 \,,
        \label{kmuPi}\\ 
    k^\nu \Pi_{\mu\nu}(k) &=& 0 \,.
    \label{knuPi}
  \end{eqnarray}
  \end{subequations}

\item [Bose symmetry~:] Bose symmetry is the statement that the
  Lagrangian should be invariant under the interchange of the two
  photon field factors.  Interchange of the two photon fields means
  the changes
  \begin{eqnarray}
    k \to -k \,, \qquad \mu \leftrightarrow \nu \,.
  \end{eqnarray}
  Clearly \Eqn{L0} satisfies this interchange trivially.  For the
  extra contribution of \Eqn{L'}, however, it implies that the
  self-energy function must satisfy the relation
  \begin{eqnarray}
    \Pi_{\mu\nu} (k) = \Pi_{\nu\mu} (-k) \,.
    \label{Bose}
  \end{eqnarray}
  So, $\Pi_{\mu\nu}$ need not be symmetric in its indices, as is often
  claimed.  \Eqn{Bose} only says that in $\Pi_{\mu\nu}$, the terms
  symmetric in the indices should be even in $k$, whereas the
  antisymmetric terms should be odd \cite{Nieves:1988qz}.

\item [Hermiticity~:] There is a constraint from the hermiticity of
  the Lagrangian.  Since the co-ordinate space version of $\lag$ must
  be hermitian, we must have
  \begin{eqnarray}
    \Big[ \lag(k) \Big]^\dagger = \lag(-k) \,.
    \label{lagkdag}
  \end{eqnarray}
  The $\lag_0$ part automatically satisfies this condition.  To see
  what it implies for $\lag'$, let us first note that in the
  co-ordinate space, the field $A^\mu(x)$ is a hermitian field, so
  that its Fourier transform satisfies the relation
  \begin{eqnarray}
    \Big( A^\mu(-k) \Big)^\dagger =  A^\mu(k^*) \,.
    \label{hermA}
  \end{eqnarray}
  In this paper, we discuss only the dispersive part of the
  self-energy for which $k$ is real.  Then, taking the hermitian
  conjugate of \Eqn{L'}, we obtain
  \begin{eqnarray}
    \Big[ \lag'(k)\Big]^\dagger
    &=& \frac12 \Big( \Pi_{\mu\nu}(k) \Big)^* \Big(
    A^\mu(-k) \Big)^\dagger \Big( A^\nu(k) \Big)^\dagger 
    \next
    &=& \frac12 \Big( \Pi_{\mu\nu}(k) \Big)^* A^\mu(k) A^\nu(-k) \,.
    \label{L'dag}
  \end{eqnarray}

  Imposing the condition of \Eqn{lagkdag} on the $\lag'$ part, we get 
  \begin{eqnarray}
    \Big( \Pi_{\mu\nu}(k) \Big)^* = \Pi_{\mu\nu}(-k) \,,
    \label{hermL'}
  \end{eqnarray}
  comparing \Eqn{L'} with \Eqn{L'dag}.

\end{description}

We can summarize the results of \Eqs{Bose} {hermL'} by writing
\begin{eqnarray}
  \Pi_{\mu\nu}(-k) = \Big( \Pi_{\mu\nu}(k) \Big)^* = \Pi_{\nu\mu}(k) \,.
  \label{hermPi}
\end{eqnarray}
Any tensor can be written as a sum of a symmetric and an antisymmetric
tensor.  \Eqn{hermPi} says that, for the dispersive part, 
\begin{enumerate}
\item The symmetric part of $\Pi_{\mu\nu}$ would be real and
  an even function of $k$.

\item The antisymmetric part of $\Pi_{\mu\nu}$ would be purely
  imaginary, and odd in $k$.

\end{enumerate}

In order to explore properties of the self-energy function, it will be
convenient to define a matrix $\mat\Pi(k)$ whose element in the
$\mu\sup{th}$ row and $\nu\sup{th}$ column is $\Pi^\mu{}_\nu(k)$.
Note that these are the objects with one up and one down index, not
the objects that appear in \Eqn{hermPi}.  The object $\Pi_{\mu\nu}$ is
the element in the $\mu\sup{th}$ row and $\nu\sup{th}$ column of the
matrix $\mat\eta\mat\Pi$, where $\mat\eta$ is a matrix whose element
in the $\mu\sup{th}$ row and $\nu\sup{th}$ column is $\eta_{\mu\nu}$.
In the matrix notation, \Eqn{hermPi} becomes
\begin{eqnarray}
  \Big[ \mat\eta \mat\Pi(-k) \Big]^\top = \Big[ \mat\eta \mat\Pi(k) 
    \Big]^\dagger = \mat\eta \mat\Pi(k) \,, 
  \label{hermetaPi}
\end{eqnarray}
It will be easier to understand these equations if we write them in
terms of the matrix elements.  For that, we only need to raise the
index $\mu$ in \Eqn{hermPi}.  Since
\begin{eqnarray}
  \Pi_\nu{}^\mu = \eta_{\nu\alpha} \Pi^\alpha{}_\beta \eta^{\beta\mu}
  \,, 
\end{eqnarray}
we obtain
\begin{eqnarray}
  \Pi^\mu{}_\nu(-k) = \Big( \Pi^\mu{}_\nu(k) \Big)^* =
  \eta_{\nu\alpha} \Pi^\alpha{}_\beta(k) \eta^{\beta\mu} \,,
\end{eqnarray}
or, more explicitly, 
\begin{subequations}
  \label{Pimunu*}
\begin{eqnarray}
  \Pi^0{}_0 (-k) = \Big[ \Pi^0{}_0 (k) \Big]^* &=& \Pi^0{}_0 (k) \,, \\
  \Pi^0{}_j (-k) = \Big[ \Pi^0{}_j (k) \Big]^* &=& - \Pi^j{}_0 (k) \,, \\
  \Pi^i{}_j (-k) = \Big[ \Pi^i{}_j (k) \Big]^* &=& \Pi^j{}_i (k) \,.
\end{eqnarray}
\end{subequations}
These equations clearly tell us which elements of the matrix $\mat\Pi$
are real and even functions of $k$, and which are imaginary and odd
functions of $k$.  The implication of these conditions on various form
factors appearing in the self-energy will be indicated when we talk
about specific backgrounds in Section\,\ref{s:sb}.

\section{Polarization vectors}\label{s:pv}
So far, we have not interpreted the gauge invariance condition in the
matrix language.  We now notice that \Eqn{knuPi} can be written as
\begin{eqnarray}
  \Pi^\mu{}_\nu k^\nu = 0 \,.
\end{eqnarray}
In the matrix notation, this equation becomes
\begin{eqnarray}
  \mat\Pi \mat k = 0 \,,
  \label{Pik=0}
\end{eqnarray}
where $\mat k$ is the column matrix whose elements are $k^\mu$.  This
is an eigenvalue equation, implying that $\mat k$ is an eigenvector of
the matrix $\mat\Pi$, and the corresponding eigenvalue is zero.

This fact prompts us to look at the eigensystem of $\mat\Pi$.  We can
define eigenvectors of this matrix through the relation
\begin{eqnarray}
  \mat\Pi \mat \epsilon_A = \Lambda_A \mat \epsilon_A \,,
  \label{eigR}
\end{eqnarray}
where the $\Lambda_A$'s are eigenvalues.  Note that, although we have
used the convention of implied summation on Lorentz indices earlier,
there is no implied sum on the index that labels the eigenvectors,
here or elsewhere in this article.  Also, it does not matter whether
this index appears as subscript or superscript: they mean the same
thing.  We just put the index wherever it is convenient in any
formula.

To be precise, \Eqn{eigR} tells us that the $\mat\epsilon_A$'s are the
right eigenvectors of the matrix $\mat\Pi$.  To check what the left
eigenvectors are, we multiply \Eqn{eigR} from the left by the matrix
$\mat\eta$ and take the hermitian conjugate of the resulting equation
to obtain
\begin{eqnarray}
  \mat\epsilon_A^\dagger \Big( \mat\eta \mat\Pi \Big)^\dagger =
  \Lambda_A^* \Big( \mat\eta \mat\epsilon_A \Big)^\dagger \,.
\end{eqnarray}
We can now use \Eqn{hermetaPi}, and write this equation in the form
\begin{eqnarray}
  \Big( \mat\eta \mat\epsilon_A \Big)^\dagger \mat\Pi = \Lambda_A^*
  \Big( \mat\eta \mat\epsilon_A \Big)^\dagger \,,
  \label{eigL}
\end{eqnarray}
using $\mat\eta=\mat\eta^\dagger$.

\Eqn{eigL} is also an eigenvalue equation, except that here the
eigenvectors multiply the matrix from the left.  We see two things
from this equation.  First, we see that the eigenvalues $\Lambda_A$
must be real, since they are defined as the solution of the equation
\begin{eqnarray}
  \det (\mat\Pi - \Lambda \mat 1) = 0 \,,
\end{eqnarray}
an equation which makes no reference to the right or left
eigenvectors.  Second, we see that, corresponding to a particular
eigenvalue $\Lambda_A$, the right eigenvector $\mat R_A$ and the left
eigenvector $\mat L_A$ are related by the matrix $\mat\eta$:
\begin{eqnarray}
  \mat R_A = \mat\epsilon_A \qquad \Rightarrow \qquad \mat L_A =
  \mat\eta \mat\epsilon_A \,.
\end{eqnarray}

It is not surprising that the right and left eigenvectors are not the
same.  In fact, this is expected for any matrix which is not normal,
i.e., which does not commute with its hermitian conjugate.  Since
$\mat\Pi$ is not a normal matrix, some properties of the left and
right eigenvectors of such matrices are worth summarizing here
\cite{nonnormal}.

Let us adopt a more general notation, and write the right and left
eigenvector equations of a matrix $\mat M$ as
\begin{subequations}
  \label{eigM}
\begin{eqnarray}
  \mat M \mat R_A &=& \mu_A \mat R_A \,, \\ 
  \mat L_B^\dagger \mat M &=& \mu_B \mat L_B^\dagger \,.
\end{eqnarray}
\end{subequations}
If we multiply the first equation by $\mat L_B^\dagger$ from the left
and the second equation by $\mat R_A$ from the right, the left sides
of the two resulting equations will be the same, implying the
following relation involving the right sides:
\begin{eqnarray}
  (\mu_A - \mu_B) \mat L_B^\dagger \mat R_A = 0 \,.
\end{eqnarray}
Thus, $\mat L_B^\dagger \mat R_A = 0$ if $\mu_A \neq \mu_B$, and we
can can choose the right and the left eigenvectors in a way that the
equation
\begin{eqnarray}
  \mat L_B^\dagger \mat R_A = \zeta_A \delta_{AB} \,,
  \label{orthogon}
\end{eqnarray}
holds, where the $\zeta_A$'s are normalization constants.

The second important property concerns the matrix defined as
\begin{eqnarray}
  \mat X = \sum_A {1 \over \zeta_A} \mat R_A \mat L_A^\dagger \,.
\end{eqnarray}
Clearly, using \Eqn {orthogon}, we see that
\begin{subequations}
  \label{XR,LX}
\begin{eqnarray}
  \mat X \mat R_B &=& \mat R_B \,, \\
  \mat L_B^\dagger \mat X &=& \mat L_B^\dagger \,.
\end{eqnarray}
\end{subequations}
This means that all eigenvalues of $\mat X$ are equal to 1, implying
that $\mat X$ is the identity matrix:
\begin{eqnarray}
  \sum_A {1 \over \zeta_A} \mat R_A \mat L_A^\dagger = \mat 1 \,.
  \label{complete}
\end{eqnarray}
A particularly useful outcome of this relation can be seen by
multiplying by $\mat M$ from the left side, and using \Eqn{eigM} to
obtain 
\begin{eqnarray}
  \mat M = \sum_A {\mu_A \over \zeta_A} \mat R_A \mat L_A^\dagger \,.
  \label{M}
\end{eqnarray}
It shows that a matrix can be written in terms of its eigenvectors and
eigenvalues.

Let us now leave the general discussion and return to the matrix $\mat
\Pi$ obtained from the self-energy.   We have already noticed that
$\mat k$ is one of its right eigenvectors.  We denote this fact by
writing 
\begin{eqnarray}
  \mat \epsilon \bsub0 \propto \mat k \,,
\end{eqnarray}
or, in component notation, 
\begin{eqnarray}
  \epsilon^\mu \bsub0 \propto k^\mu \,.
\end{eqnarray}

Leaving this one aside, there are three more eigenvectors.  These are
the polarization vectors, which we will denote by $\mat\epsilon_a$,
with a lowercase roman index that runs from 1 to 3.  The corresponding
left eigenvectors will be $\mat\eta\mat\epsilon_a$.

\Eqn{orthogon} now implies the relation
\begin{eqnarray}
   \mat k^\dagger \mat\eta \mat\epsilon_a = 0 \,.
\end{eqnarray}
In indexed notation, this reads
\begin{eqnarray}
  k^\mu \epsilon_\mu^a = 0 \,.
  \label{keps}
\end{eqnarray}
There is also the relation of orthogonality of the polarization
vectors.  If the momentum vector $k^\mu$ is a timelike vector,
the polarization vectors must be spacelike, and we can impose
\begin{eqnarray}
  \mat \epsilon_a^\dagger \mat\eta \mat\epsilon_b = - \delta_{ab} \,,
  \label{edage}
\end{eqnarray}
which amounts to the choice 
\begin{eqnarray}
  \zeta_a = -1 \quad \mbox{for $a=1,2,3$}.
\end{eqnarray}

In indexed notation, \Eqn{edage} means
\begin{eqnarray}
  \Big( \epsilon_a^\mu \Big)^* \epsilon^b_\mu = - \delta_{ab} \,,
\end{eqnarray}
and \Eqn{M} gives
\begin{eqnarray}
  \mat\Pi = - \sum_a \Lambda_a \mat\epsilon_a \Big(
  \mat\eta\mat\epsilon_a \Big)^\dagger \,. 
\end{eqnarray}
The $\Lambda_a$'s are the eigenvalues of $\mat\Pi$, as defined
earlier.  Note that the sum is only on the three polarization vectors.
The other eigenvector does not contribute since it is associated with
a null eigenvalue.  In the indexed notation, we can write the last
equation as
\begin{eqnarray}
  \Pi^\mu{}_\nu = - \sum_a \Lambda_a \epsilon_a^\mu
  (\epsilon^a_\nu)^* \,.
\end{eqnarray}

This relation is easier to deal with using the matrix
$\mat\eta\mat\Pi$ which has both lower indices.  We write
\begin{eqnarray}
  \Pi_{\rho\nu} = \sum_a \Lambda_a P^a_{\rho\nu} \,,
  \label{LamP}
\end{eqnarray}
where
\begin{eqnarray}
  P^a_{\rho\nu} = - \epsilon^a_\rho (\epsilon^a_\nu)^* \,.
  \label{P}
\end{eqnarray}
Note that these objects satisfy the relation
\begin{eqnarray}
  \eta^{\nu\lambda} P^a_{\mu\nu} P^b_{\lambda\rho} = \delta_{ab} 
  P^a_{\mu\rho} \,. 
\end{eqnarray}
So the $P_a$'s are, in some sense, projectors for the different
polarization vectors.  Because of \Eqn{keps}, they satisfy the
relations
\begin{subequations}
  \label{kP}
\begin{eqnarray}
  k^\mu P^a_{\mu\nu} &=& 0 \,, \\ 
  k^\nu P^a_{\mu\nu} &=& 0 \,.
\end{eqnarray}
\end{subequations}
Note that these equations are exactly similar to those in \Eqn{kPi}.
This must be the case since $\Pi_{\mu\nu}$ is a linear superposition of
these projection tensors, as shown in \Eqn{LamP}.

\Eqn{complete} provides more information about the polarization
vectors.  Applying it on the matrix $\mat\Pi$, we obtain the relation 
\begin{eqnarray}
  {k^\mu k_\nu \over k^2} - \sum_a \epsilon_a^\mu (\epsilon^a_\nu)^* =
  \delta^\mu_\nu \,,
\end{eqnarray}
which can be rewritten as
\begin{eqnarray}
  \sum_a P^a_{\mu\nu} = \tilde\eta_{\mu\nu} \,.
  \label{sumP}
\end{eqnarray}
This is the polarization sum formula.

\section{Propagator and dispersion relations}\label{s:dr}
The equation of motion of the photon field that follows from
\Eqn{L0+L'} is
\begin{eqnarray}
  \Big[ {-} k^2 \tilde\eta_{\mu\nu} + \Pi_{\mu\nu}(k) \Big] A^\nu(k) =
  j_\mu (k) \,, 
\end{eqnarray}
where $j^\mu$ is the current that the photon couples to.  The equation
for the propagator, after adding a gauge fixing term, is

\begin{eqnarray}
  \Big[  {-} k^2 \tilde\eta_{\mu\nu} + \Pi_{\mu\nu}(k) + \frac1\xi\,
    {k_\mu k_\nu \over k^2} \Big] D^{\nu\rho} = \delta^\rho_\mu \,.
  \label{eqmotion}
\end{eqnarray}
Substituting $\tilde\eta_{\mu\nu}$ from \Eqn{sumP} and using the
expression of \Eqn{LamP} for $\Pi_{\mu\nu}$, we obtain the following
equation that defines the propagator:
\begin{eqnarray}
  \Big[ \sum_a \Big( {-} k^2 + \Lambda_a \Big) P^a_{\mu\nu}
    + \frac1\xi \, {k_\mu k_\nu \over k^2} \Big] D^{\nu\rho} =
  \delta^\rho_\mu \,.
  \label{eqprop}
\end{eqnarray}
It can now be easily seen that the propagator is given by 
\begin{eqnarray}
  D^{\mu\nu} (k) &=& - \sum_a {P_a^{\mu\nu} \over
    k^2 - \Lambda_a} + {\xi \over k^2} \, {k^\mu k^\nu \over k^2}
  \next 
  &=& \sum_a {\epsilon_a^\mu {\epsilon_a^\nu}^*  \over
    k^2 - \Lambda_a} + {\xi \over k^2} \, {k^\mu k^\nu \over k^2}
  \,. 
  \label{prop}
\end{eqnarray}
For the polarization $a$, the dispersion relation is the relation for
which the propagator blows up, i.e.,
\begin{eqnarray}
  k^2 = \Lambda_a \,.
  \label{disp}
\end{eqnarray}
It has to be remembered that $\Lambda_a$ is a function of the
momentum.  Thus, this is an implicit equation that has to be solved to
obtain the dispersion relation.  The polarization vector that
satisfies this dispersion relation is $\epsilon^\mu_a$.

\section{Examples of specific backgrounds}\label{s:sb}
We will now illustrate how this formalism applies to various
non-trivial backgrounds.  We will always choose the photon momentum in
the $x$-direction, i.e., in matrix form we will have
\begin{eqnarray}
  \mat k = \left( \begin{array}{c} \omega \\ K \\ 0 \\ 0
  \end{array} \right) \,.
  \label{matk}
\end{eqnarray}
Then the components of the matrix $\tilde{\mat\eta}$ are given by
\begin{eqnarray}
  \tilde{\mat\eta} = {1 \over k^2}
  \left( \begin{array}{cccc}
    -K^2 & \omega K & 0 & 0 \\
    -\omega K & \omega^2 & 0 & 0 \\
    0 & 0 & k^2 & 0 \\
    0 & 0 & 0 & k^2 \\ 
  \end{array}  \right) \,,
\end{eqnarray}
remembering that the matrix form corresponds to the components of the
mixed tensor, whose first index is contravariant and the second
index is covariant.  Other tensors necessary for building up the
self-energy tensor depend on the type of background, and will be
defined as we go along.

\subsection{Thermal background}\label{s:sb:th}
A thermal background is characterized by a temperature and a chemical
potential, both of which are scalars.  There is one vector associated
with a thermal medium, which is the center of mass velocity of the
medium.  Let us call it $u^\mu$.  We can define
\begin{eqnarray}
  \tilde u_\mu = \tilde \eta_{\mu\rho} u^\rho \,,
\end{eqnarray}
which satisfies the relation
\begin{eqnarray}
  k^\mu \tilde u_\mu = 0 \,.
  \label{kutil}
\end{eqnarray}
Using this, one can form the tensor
\begin{eqnarray}
  L_{\mu\nu} = {\tilde u_\mu \tilde u_\nu \over \tilde u^2} \,,
\end{eqnarray}
which vanishes when contracted with either $k^\mu$ or $k^\nu$.
Therefore, it is a tensor that can be used for writing $\Pi_{\mu\nu}$.
Of course $\tilde\eta_{\mu\nu}$ is another such tensor.  So, we can
write the self-energy as~\cite{Kislinger:1975uy}
\begin{eqnarray}
  \Pi_{\mu\nu} = a \tilde\eta_{\mu\nu} + b L_{\mu\nu} \,,
  \label{th.Pi}
\end{eqnarray}
where $a$ and $b$ are Lorentz invariants.  

The matrix $\mat\Pi$ follows from \Eqn{th.Pi} once we decide on the
components of $u^\mu$.  Let us work in a frame where the medium is at
rest, so that the time component of $u^\mu$ is equal to 1 and all
other components are zero.  In this frame, we find
\begin{eqnarray}
  \mat\Pi = {1 \over k^2} \left( \begin{array}{cccc}
    -(a + b) K^2 & (a + b) \omega K & 0 & 0 \\ 
    -(a + b) \omega K & (a + b) \omega^2 & 0 & 0 \\ 
    0 & 0 & ak^2 & 0 \\
    0 & 0 & 0 & ak^2 \\
  \end{array}  \right) \,.
  \label{th.matPi}
\end{eqnarray}
\Eqn{Pimunu*} implies that the form factors $a$ and $b$ are real, and
are even functions of $k$.

It is easy to check that $\mat k$, defined in \Eqn{matk}, is a right
eigenvector and $\mat\eta\mat k$ is a left eigenvector of this matrix,
with eigenvalue equal to zero.  Among the others, there is a
non-degenerate eigenvalue
\begin{eqnarray}
  \Lambda \bsub L = a+b \,,
\end{eqnarray}
whose eigenvector is $\tilde u^\mu$.  In matrix notation, we can write
the eigenvector corresponding to this mode as
\begin{eqnarray}
  \mat\epsilon \bsub L \propto \tilde{\mat u} \,, 
\end{eqnarray}
and we can define the projector corresponding to this mode as
\begin{eqnarray}
  P \bsup L_{\mu\nu} = L_{\mu\nu} \,.
\end{eqnarray}
Since the spatial component of this polarization vector is along the
direction of the photon momentum, this is called the longitudinal
mode.

In compliance with \Eqn{sumP}, the sum of the projectors of the other
two modes will be given by
\begin{eqnarray}
  P \bsup T_{\mu\nu} = \tilde\eta_{\mu\nu} - P \bsup L_{\mu\nu} \,,
\end{eqnarray}
and these two modes will be transverse and degenerate, with
eigenvalues
\begin{eqnarray}
  \Lambda \bsub T = a \,.
\end{eqnarray}
Following \Eqn{prop}, we can now write down the photon propagator:
\begin{eqnarray}
  D^{\mu\nu} (k) &=& - {P \bsub T^{\mu\nu} \over
    k^2 - \Lambda \bsub T} - {P \bsub L^{\mu\nu} \over
    k^2 - \Lambda \bsub L} + {\xi \over k^2} \, {k^\mu k^\nu \over
    k^2} \,.
\end{eqnarray}
The dispersion relations of the different modes follow from \Eqn{disp}.

\subsection{Chiral thermal medium}
It was pointed out \cite{Nieves:1988qz, Nieves:1992et} that
\Eqn{th.Pi} is not the most general self-energy tensor that one can
write using the two 4-vectors $k^\mu$ and $u^\mu$.  Rather, the
general form would be
\begin{eqnarray}
  \Pi_{\mu\nu} = a \tilde\eta_{\mu\nu} + b L_{\mu\nu} + ic C_{\mu\nu}
  \,, 
  \label{ch.Pi}
\end{eqnarray}
where
\begin{eqnarray}
  C_{\mu\nu} = {1 \over \kappa} \, \varepsilon_{\mu\nu\lambda\rho}
  k^\lambda u^\rho \,.
  \label{ch.PC}
\end{eqnarray}
where 
\begin{eqnarray}
  \kappa = \Big| \sqrt{(k\cdot u)^2 - k^2} \Big| \,, 
  \label{kappa}
\end{eqnarray}
a Lorentz invariant factor inserted to ensure that $C_{\mu\nu}$ is
dimensionless, at par with the other tensors that appear in
\Eqn{ch.Pi}.  With our choice of \Eqn{matk}, $\kappa = |K|$.

Note that the extra term in the self-energy is not symmetric in the
Lorentz indices.  But this symmetry was never one of the requirements
mentioned in Sec.~\ref{s:se}.  The factor of $i$ accompanying this
tensor in \Eqn{ch.Pi} ensures that the associated form factor $c$ is
real, according to \Eqn{hermL'}.

The matrix $\mat\Pi$ can now be easily constructed, and it is
\begin{eqnarray}
  \mat\Pi = {1 \over k^2} \left( \begin{array}{cccc}
    -(a + b) K^2 & (a + b) \omega K & 0 & 0 \\ 
    -(a + b) \omega K & (a + b) \omega^2 & 0 & 0 \\ 
    0 & 0 & ak^2 & ick^2 K/\kappa \\
    0 & 0 & -ick^2 K/\kappa & ak^2 \\
  \end{array}  \right) \,.
\end{eqnarray}
\Eqn{Pimunu*} shows that all three form factors, $a$, $b$ and $c$, are
even functions of $k$, and all are real.

The longitudinal eigenvector and its eigenvalue are exactly the same as
that found in the earlier case.  But the transverse eigenvalues are no
more degenerate.  They, and their corresponding eigenvectors, are as
follows:
\begin{eqnarray}
  \Lambda \bsub \pm = a \pm c K/\kappa, &  \qquad
  \mat\epsilon \bsub \pm \propto & \left( \begin{array}{c} 0 \\ 0 \\ 1
    \\ \mp i 
  \end{array} \right) \,.
\end{eqnarray}
Thus, the propagating modes are circularly polarized modes, and the
left and right circular polarized waves have different dispersion
relations~\cite{Nieves:1988qz, Nieves:1992et}.  The linearly polarized
transverse waves are not eigenvectors of propagation.  If one sends in
a linearly polarized wave, its direction of polarization will rotate.
This is the phenomenon of optical activity.

\subsection{Background magnetic field}\label{s:sb:mf}
Optical activity can also be induced by a background magnetic field.
This phenomenon is called the Faraday effect.  In order to investigate
it, we first need to find tensors built from the background field
tensor $B_{\mu\nu}$ whose contraction vanishes with the photon
momentum vector.  Two such tensors were identified
\cite{Ganguly:1999ts}: 
\begin{eqnarray}
  M_{\mu\nu} &=& {1 \over k^2} \, \varepsilon_{\mu\nu\sigma\tau}
  k^\sigma k_\lambda B^{\lambda\tau} \,,  \next 
  M'_{\mu\nu} &=& B_{\mu\nu} - {k_\mu k^\lambda
    B_{\lambda\nu} \over k^2} + {k_\nu k^\lambda
    B_{\lambda\mu} \over k^2} \,.
  \label{MM'}
\end{eqnarray}
In this article, just to keep the formulas simple-looking, we will
assume that the self-energy tensor does not have a term proportional
to $M'_{\mu\nu}$, i.e., we have
\begin{eqnarray}
  \Pi_{\mu\nu} = a \tilde\eta_{\mu\nu} + ib M_{\mu\nu} \,.  
  \label{mf.Pi}
\end{eqnarray}
Note that we have put a factor of $i$ in the second term, in order to
make sure that the form factor $b$ is real, as demanded from
\Eqn{hermPi}.    It is also to be noted that the form factor $b$ has
to be an odd function of $k$, which follows from \Eqn{Bose}.

In order to write the matrix $\mat \Pi$ explicitly, 
let us say that the background magnetic field is in the $x$-$y$ plane,
making an angle $\beta$ with the $x$-axis.  Then the only non-zero
components of the background field tensor $B_{\mu\nu}$ are the
following:
\begin{eqnarray}
  B_{23} = - B_{32} &=& \cos\beta \,, \next 
  B_{31} = - B_{13} &=& \sin\beta \,, 
  \label{B}
\end{eqnarray}
where we have normalized the magnitude of the magnetic field to be
unity.  The matrix $\mat \Pi$ that follows from \Eqn{mf.Pi} is then
easily calculated:
\begin{eqnarray}
  \mat\Pi = {1 \over k^2} \left( \begin{array}{cccc}
    -aK^2 & a \omega K & -ib' K^2 & 0 \\ 
    -a \omega K & a \omega^2 & -ib' \omega K & 0 \\ 
    -ib'K^2 & ib' \omega K & ak^2 & 0 \\
    0 & 0 & 0 & ak^2 \\
  \end{array}  \right) \,,
\end{eqnarray}
where
\begin{eqnarray}
  b' = b \sin \beta \,.
\end{eqnarray}
Note that if the magnetic field is parallel to the photon momentum,
i.e., if $\sin\beta=0$, there is no effect.  For what follows, we
assume that $b'\neq0$, i.e., the magnetic field is not parallel to the
photon momentum.

Apart from the trivial eigenvector which is the photon momentum, we
can easily see that there is one eigenmode with
\begin{eqnarray}
  \Lambda \bsub 3 = a, &  \qquad
  \mat\epsilon \bsub 3 = & \left( \begin{array}{c} 0 \\ 0 \\ 0 \\ 1
  \end{array} \right) \,.
  \label{eps3}
\end{eqnarray}
This is the mode that is perpendicular to both the magnetic field and
the photon 3-momentum.  The other two eigenmodes can be easily solved,
with the result:
\begin{eqnarray}
  \Lambda \bsub \pm = a \pm {b'K \over \sqrt{k^2}}, &  \qquad
  \mat\epsilon \bsub \pm = & \left( \begin{array}{c} K \\ \omega \\
    \pm i \sqrt{k^2} \\ 0 
  \end{array} \right) \,.
  \label{sbmf.eig+-}
\end{eqnarray}
These are in general elliptically polarized states.  As in the case of
chiral thermal medium, we see that the linearly polarized states are
not eigenmodes of propagation.  Thus, if a linearly polarized wave is
sent through a magnetic field, its polarization vector will rotate
as it moves.  This is Faraday rotation~\cite{Ganguly:1999ts}.

\subsection{Magnetic field in a thermal medium}\label{s:sb:mm}
We now go to the more complicated case where there is a thermal bath
as well as a magnetic field.  Because of the presence of the 4-vector
$u^\mu$, we can define now \cite{Karmakar:2018aig} a 4-vector for the
magnetic field by the relation
\begin{eqnarray}
  n^\mu = {1 \over 2} \varepsilon^{\mu\nu\lambda\rho} u_\nu
  B_{\lambda\rho} \,.
\end{eqnarray}
It is easier to work with this vector rather than with the background
field tensor.

We can define an associated vector
\begin{eqnarray}
  \tilde n_\mu = \tilde \eta_{\mu\nu} n^\nu \,,
\end{eqnarray}
which satisfies the relation
\begin{eqnarray}
  k^\mu \tilde n_\mu = 0 \,.
\end{eqnarray}
So, in $\Pi_{\mu\nu}$, we can have a term proportional to
\begin{eqnarray}
  n_{\mu\nu} = {\tilde n_\mu \tilde n_\nu \over \tilde n^2} \,,
\end{eqnarray}
because the said term will satisfy \Eqn{kPi}.  There are other tensors
which will satisfy \Eqn{kPi}.  For example, one can consider
\begin{eqnarray}
  S_{\mu\nu} &=& {\tilde u_\mu \tilde n_\nu + \tilde n_\mu \tilde
    u_\nu \over \sqrt{\tilde u^2 \, \tilde n^2}} \,.
  \label{S}
\end{eqnarray}
Detailed analysis of photon dispersion was made
\cite{Karmakar:2018aig} with these four tensors.  However, there can
be many other tensors, including some which are antisymmetric in the
indices.  The list is quite long \cite{Ganguly:1999ts}.  Here, we will
take one of the antisymmetric tensors for the purpose of illustration: 
\begin{eqnarray}
  D_{\mu\nu} &=& {\tilde u_\mu \tilde n_\nu - \tilde n_\mu \tilde
    u_\nu \over \sqrt{\tilde u^2 \, \tilde n^2}} 
  \,. \label{D}
\end{eqnarray}
So we will define the self-energy tensor as a sum involving these 
tensors:
\begin{eqnarray}
  \Pi_{\mu\nu} = a \tilde\eta_{\mu\nu} +  b L_{\mu\nu} + c 
  n_{\mu\nu} + d S_{\mu\nu} + i d' D_{\mu\nu} \,.
  \label{mm.Pi}
\end{eqnarray}

With the velocity 4-vector of the thermal bath defined as in
Sec.~\ref{s:sb:th}, and the magnetic field in this frame assumed to
have the components specified in Sec.~\ref{s:sb:mf}, we can construct
the matrix $\mat\Pi$ easily, using \Eqn{mm.Pi}.  It will be given by
\begin{eqnarray}
  \mat\Pi = {1 \over k^2} \left( \begin{array}{cccc}
    -\sigma K^2 & \sigma \omega K & \gamma {K \over \omega} k^2 & 0 \\
    -\sigma \omega K & \sigma \omega^2 & \gamma k^2 & 0 \\
    - \gamma^* {K \over \omega} k^2 & \gamma^* k^2 & \sigma' k^2 & 0 \\  
    0 & 0 & 0 & ak^2 \\
  \end{array}  \right) \,.
\end{eqnarray}
where we have used the shorthands
\begin{subequations}
  \label{sbmm.shorthands}
\begin{eqnarray}
  \sigma &=& a + b + c \cos^2\alpha + 2d \cos\alpha \,, \\
  \sigma' &=& a + c \sin^2\alpha \,, \\
  \gamma &=& \Big[ c \cos\alpha + (d+id') \Big] \tan\beta \cos\alpha \,.
\end{eqnarray}
\end{subequations}
The angle $\alpha$ appearing in these equations is related to the
angle $\beta$ that appears in \Eqn{B} through the formula
\begin{eqnarray}
  \cos\alpha = {\omega \cos\beta \over \displaystyle\sqrt{\omega^2
      \cos^2\beta + k^2 \sin^2\beta}} \,.
  \label{sbmm.alpha}
\end{eqnarray}
One eigenvalue and the corresponding eigenvector is exactly the same
as that given in \Eqn{eps3}.  The other two non-zero eigenvalues
satisfy the equation
\begin{eqnarray}
  \Lambda^2 - (\sigma + \sigma') \Lambda + \Big( \sigma \sigma' -
  |\gamma|^2 {k^2 \over \omega^2} \Big) = 0 \,,
\end{eqnarray}
so that
\begin{eqnarray}
  \Lambda = \frac12 \Big[ \sigma + \sigma' \pm \sqrt{(\sigma -
      \sigma')^2 + 4 |\gamma^2| k^2 / \omega^2} \Big] \,.
  \label{sbmm.roots}
\end{eqnarray}

The eigenvectors corresponding to these two eigenvalues are of the form
\begin{eqnarray}
  \mat\epsilon \bsub a \propto \left( \begin{array}{c} \gamma K
    \\ \gamma \omega
    \\ {(\Lambda \bsub a - \sigma) \omega} \\ 0 
  \end{array} \right) \qquad \mbox{for $a=1,2$}.
  \label{mm.evec}
\end{eqnarray}
In fact, the results of both \Sect{s:sb:th} and \Sect{s:sb:mf} can be
seen as special cases of the result given here.

\section{Axis-free notation}\label{s:af}
Throughout \Sect{s:sb}, we have made specific choices for the
directions of the photon momentum and other vectors such as $u^\mu$
and $b^\mu$.  This fact should not be seen as a limitation of the
method described.  Often, the choices meant no loss of generality.
For example, consider the choices made in \Sect{s:sb:mf}.  We can
always choose the axes in such a way that the 3-momentum of the photon
is in the $x$-direction, and then define the $x$-$y$ plane to contain
the direction of the magnetic field.  There is no loss of generality
in assuming the forms given in \Eqs{matk} {B}.

If one wants different directions for some reason, one just needs to
write the relevant vectors in the appropriate form.  For example, if
one wants to keep the direction of the photon 3-momentum arbitrary in
the co-ordinate axes, one just needs to redo the exercise by replacing
\Eqn{matk} with a different matrix, whose last three elements should
be the Cartesian components of a 3-vector in terms of the magnitude
$\kappa$ and the polar and azimuthal angles.  It will just make the
matrices look a bit more cumbersome, but the task remains the same in
principle.

However, it is not difficult to take an alternative route, viz., to
write the results in a form that is free from the choices of the axes.
All we need to do is to recast the equations in terms of the vectors
and tensors that appear in the problem.  This was already done for the
case of a thermal background, where we saw that the longitudinal
polarization vector is just $\tilde u^\mu$.  Here, we outline how such
a treatment can be extended to problems with a magnetic field.

The important task is to find a set of four vectors which are linearly
independent, and can therefore serve as a basis.  One of them is
surely $k^\mu$, and we define the notation
\begin{eqnarray}
  e^\mu \bsub 0 = k^\mu \,.
  \label{af.e0}
\end{eqnarray}
There is no pressing need for normalizing any of the basis vectors, so
we won't bother.

\subsection{Background magnetic field}\label{s:af:mf}
Here, we need to construct some vectors which are orthogonal to
$k^\mu$.  We can construct a series of vectors, contracting $k^\mu$
with different powers of the magnetic field tensor, like
\begin{eqnarray}
  (k\cdot B)^\mu &=& k_\lambda B^{\lambda\mu} \,, \next
  (k \cdot B \cdot B)^\mu &=& (k\cdot B)_\lambda B^{\lambda\mu} =
  k^\rho B_{\rho\lambda} B^{\lambda\mu} \,, 
  \label{afmf.kB}
\end{eqnarray}
and so on.  We can use these vectors to form a mutually orthogonal set
of vectors.  Starting with $k^\mu$, we can apply the Gram-Schmidt
orthogonalization process to obtain the next two in the set:
\begin{subequations}
  \label{afmf.basis}
\begin{eqnarray}
  e^\mu \bsub 1 &=& k_\lambda B^{\lambda\mu} \equiv (k\cdot B)^\mu \,,
  \\
  e^\mu \bsub 2 &=& (k \cdot B \cdot B)^\mu + {k^\mu \over k^2} (k
  \cdot B)^2 \,.
\end{eqnarray}
We can go on with the next member of the type shown in \Eqn{afmf.kB}.
But it is much easier to complete the set by introducing the vector 
\begin{eqnarray}
  e^\mu \bsub3 = {1 \over k^2} \varepsilon^{\mu\nu\lambda\rho} k_\nu
  (k \cdot B)_\lambda (k \cdot B \cdot B)_\rho \,,
  \label{afmf.basis3}
\end{eqnarray}
\end{subequations}
because for any three 4-vectors $A^\mu$, $B^\mu$, $C^\mu$, the object
\begin{eqnarray}
  V^\mu = \varepsilon^{\mu\nu\lambda\rho} A_\nu B_\lambda C_\rho
  \label{ABC}
\end{eqnarray}
will be orthogonal to all three.  The prefactor $1/k^2$ has been put
in \Eqn{afmf.basis3} just to ensure that $e^\mu
\bsub3$ has the same physical dimension as the other two.  It is not
really necessary to ensure that, but it is convenient.

We can now see the effect of contracting the self-energy tensor, given
in \Eqn{mf.Pi}, with these basis vectors.  Contraction with $k^\mu$
will give zero, of course.  For the other ones, we get
\begin{subequations}
  \label{afmf.Pie} 
\begin{eqnarray}
  \Pi^\mu{}_\nu e^\nu \bsub1 &=& a e^\mu \bsub1 \,,
  \label{afmf.Pie1} \\
  \Pi^\mu{}_\nu e^\nu \bsub2 &=& a e^\mu \bsub2 + ib e^\mu \bsub3 \,, 
  \label{afmf.Pie2} \\
  \Pi^\mu{}_\nu e^\nu \bsub3 &=& a e^\mu \bsub3 + ib {(k \cdot B)^2
    \over k^2} e^\mu \bsub2 \,.
  \label{afmf.Pie3} 
\end{eqnarray}
\end{subequations}
\Eqn{afmf.Pie1} shows that $e^\mu \bsub1$ is an eigenvector, with
eigenvalue $a$.  For the choice of the axes made in \Sect{s:sb:mf},
this eigenvector happened to be along the $z$-axis, which is what
shows in \Eqn{eps3}.  The other two eigenvectors are linear
combinations of the basis vectors $e^\mu \bsub2$ and $e^\mu \bsub3$.
These eigenvectors can easily be found, and the result is:
\begin{eqnarray}
  \Lambda \bsub\pm = a \pm rb , &  \qquad
  \mat\epsilon \bsub \pm = & \mat e \bsub 3 \pm ir  \mat e
  \bsub 2 \,,
  \label{afmf.eig+-}
\end{eqnarray}
where
\begin{eqnarray}
  r^2 = - {(k \cdot B)^2 \over k^2} \,.
  \label{afmf.r}
\end{eqnarray}
Despite the negative sign in this formula, $r$ is a real number,
because $k\cdot B$ is a spacelike vector whereas $k$ is timelike.  We
have not made any effort for writing the eigenvectors in normalized
form.

\Eqs{afmf.Pie1} {afmf.eig+-} give the axis-free definition of the
eigenvalues and eigenvectors.  With the choices made in
\Sect{s:sb:mf}, we have $(k\cdot B)^2= -K^2\sin^2\beta$, and so the
eigenvalues shown in \Eqn{sbmf.eig+-} follow.  The eigenvectors also
reduce to the ones showed there, apart from the overall normalization
which has not been adjusted in \Eqn{afmf.eig+-} anyway.

\subsection{Magnetic field in a thermal medium}\label{s:af:mm}
Here also, we can start with the vector $k^\mu$.  Then, in the list,
we can add
\begin{subequations}
  \label{afmm.basis}
\begin{eqnarray}
  e^\mu \bsub 1 = \tilde u^\mu \,,
  \label{afmm.basis1}
\end{eqnarray}
which is orthogonal to $k^\mu$.  The next one can involve $\tilde
n^\nu$.  In order that it is orthogonal to both $e^\mu \bsub 0$ and
$e^\mu \bsub 1$, we choose
\begin{eqnarray}
  e^\mu \bsub 2 = \tilde n^\mu - {\tilde u \cdot \tilde n \over \tilde
  u^2} \tilde u^\mu \,.
  \label{afmm.basis2}
\end{eqnarray}
And the final one can be chosen as
\begin{eqnarray}
  \epsilon^\mu \bsub3 = \varepsilon^{\mu\nu\lambda\rho} k_\nu
  \tilde u_\lambda \tilde n_\rho = \varepsilon^{\mu\nu\lambda\rho} k_\nu
  u_\lambda n_\rho \,,
  \label{afmm.basis3}
\end{eqnarray}
\end{subequations}
which is obviously orthogonal to the other ones defined earlier.  We
do not care about the normalization at this stage, because it is not
necessary.

We now contract the self-energy tensor with these basis vectors.  It
will of course show that $k^\mu$ is an eigenvector with zero
eigenvalue.  On the other three, first of all, we will find one which
stands out, viz., 
\begin{eqnarray}
  \Pi^\mu{}_\nu e^\nu \bsub3 &=&  a  e^\mu \bsub3 \,.
  \label{afmm.Pie3}
\end{eqnarray}
It means that $e^\nu \bsub3$ is an eigenvector of the matrix
$\Pi^\mu{}_\nu$, with eigenvalue $a$.  This was the result shown in
\Eqn{eps3}.

For the other two basis vectors, we get equations of the form
\begin{eqnarray}
  \Pi^\mu{}_\nu e^\nu \bsub1 &=& P_1 e^\mu \bsub1 + Q_1 e^\mu \bsub2
  \,, \next
  \Pi^\mu{}_\nu e^\nu \bsub2 &=& P_2 e^\mu \bsub1 + Q_2 e^\mu \bsub2
  \,, 
\end{eqnarray}
where
\begin{eqnarray}
  P_1 &=& a + b + c {(\tilde u \cdot \tilde n)^2 \over \tilde u^2
    \tilde n^2} + 2d {\tilde u \cdot \tilde n \over \sqrt{\tilde u^2
      \tilde n^2}} \,, \next
  P_2 &=& {\tilde u^2 \tilde n^2 - (\tilde u \cdot \tilde n)^2 \over
    \tilde u^4 \tilde n^2} \left( c \tilde u \cdot \tilde n +
  (d+id') \sqrt{\tilde u^2 \tilde n^2} \right)
  \,, \next 
  Q_1 &=& {1 \over \tilde n^2} \left( c \tilde u \cdot \tilde n + (d-id')
  \sqrt{\tilde u^2 \tilde n^2} \right) \,, \next
  Q_2 &=& a + c \Big( 1 - {(\tilde u \cdot \tilde n)^2 \over \tilde u^2
    \tilde n^2} \Big) \,.
  \label{P1Q1P2Q2}
\end{eqnarray}
This means that the eigenvectors will be linear combinations of $e^\mu
\bsub1$ and $ e^\mu \bsub2$.  If we ignore the task of normalizing
the eigenvectors and call an eigenvector $e^\mu \bsub1+r e^\mu
\bsub2$, then $r$ will be given by
\begin{eqnarray}
  r = {1 \over 2P_2} \Big( Q_2 - P_1 \pm \sqrt{(Q_2-P_1)^2 + 4 Q_1P_2}
  \Big) \,, 
\end{eqnarray}
and the eigenvalue will be given by
\begin{eqnarray}
  \Lambda = \frac12 \Big[ Q_2 + P_1 \pm \sqrt{(Q_2-P_1)^2 + 4 Q_1P_2}
  \Big] \,.
\end{eqnarray}
It is easy to see that, with the choice of axes made in
Section\,\ref{s:sb}, this expression reduces to that in
\Eqn{sbmm.roots}.

\section{Comments}\label{s:co}
We started this article by saying that we want to proceed, as much as
possible, without committing ourselves to any specific background.  We
laid down our formalism in \Sect{s:se} through \Sect{s:dr}, where we
showed that the problem can be cast in the form of a problem in matrix
algebra.  In \Sect{s:sb}, we illustrated the method with some choices
of background and choices of specific axes.  In \Sect{s:af}, we went
one step further and found the polarization vectors and the dispersion
relation in a completely axis-free notation in the cases of a
background magnetic field, with or without a background medium.

Of course in the process one has to identify some tensors which are
orthogonal to the photon 4-momentum.  This part is the same as that in
a conventional treatment \cite{Ganguly:1999ts, Karmakar:2018aig,
  Kislinger:1975uy, Weldon:1982aq, Nieves:1988qz}.  The difference in
our approach is that, because we identify the polarization vectors as
eigenvectors of a matrix, it is easier to find the polarization
vectors.  Our approach also shows that, no matter how many tensors
participate in the making of the self-energy, it is guaranteed that
one will obtain three polarization vectors, as is expected because the
photon is a spin-1 particle.

There is another piece of information that is easily derived from the
approach of \Sect{s:af}.  This is the question of degeneracy of the
dispersion relations.  This question is related to the possibility of
defining the basis vectors in the matrix formulation.  We can always
take the photon momentum $k^\mu$ as one of the basis vectors.  If the
background is non-trivial, we will be able to add at least one more
vector to this list.  That is not enough for lifting degeneracy of the
dispersion modes.  If we have a third vector, then we can also find a
fourth one through the prescription of \Eqn{ABC}.  But even this is
not enough: the four vectors so defined must be linearly independent
in order that they can form a basis.

There is an easy check for linear independence.  In the demonstrative
examples, we always chose the fourth basis vector through \Eqn{ABC}.
Clearly, this fourth vector cannot be linearly dependent on the other
three.  We therefore need to check whether the first three basis
vectors are linearly independent.  If they are dependent, then the
vector $V^\mu$, defined in \Eqn{ABC}, will be the null vector.

Clearly, $e^\mu \bsub0$ cannot be proportional to either $e^\mu
\bsub1$ and $e^\mu \bsub2$, because the former is a timelike vector
whereas the latter ones are spacelike.  Thus, our task reduces to
finding whether any combination of $e^\mu \bsub1$ and $e^\mu \bsub2$
can be the null vector.  Either any of them will have to be the null
vector by itself, or they must be proportional to each other.

For the case dealt with in \Sect{s:af:mf}, one of these possibilities
mean that $k\cdot B$ is itself a null vector.  This means that the
magnetic field is parallel to the spatial direction of the photon
momentum vector.  In the notation of \Sect{s:sb:mf}, this means that
$\beta=0$, and definitely it shows degenerate modes.  In the axis-free
notation as well, we see that this state of affairs imply $r=0$ in
\Eqn{afmf.r}, so that we have degenerate modes.  Another alternative,
viz.\ the vanishing of $e^\mu \bsub2$, is impossible because that
requires $k \cdot B \cdot B$, a spacelike vector, to be proportional
to $k^\mu$, a timelike vector.

For the case described in \Sect{s:af:mm}, the basis vectors are
linearly dependent if $\tilde u^\mu$ and $\tilde n^\mu$ are
proportional.  In the notation of \Sect{s:sb:mm}, this means that the
angle $\beta$ vanishes, and therefore so does $\alpha$ through
\Eqn{sbmm.alpha}.  Then, from \Eqn{sbmm.shorthands}, we find that
$\gamma=0$, which means that the eigenvalues obtained from
\Eqn{sbmm.roots} are $\sigma$ and $\sigma'$.  But, for $\alpha=0$, we
get $\sigma'=a$ from \Eqn{sbmm.shorthands}, so that this root is
degenerate with the root noted down in \Eqn{eps3}.

Overall, we find that the matrix formulation of the photon propagation
problem clears up many aspects of the polarization vectors and
dispersion relations which are otherwise not easy to understand from
the usual approach.  And, although we used the photon all along to
describe the methods, they are applicable to gluons, or any other
spin-1 boson for which there is an associated gauge invariance.

\paragraph*{Acknowledgements~:} I have had numerous discussions with
Munshi Golam Mustafa and Bithika Karmakar during the time they had
been working on Ref.\,\bcite{Karmakar:2018aig}.  Both of them have
also kindly read an earlier version of this article and made helpful
comments.  At the time of working on the present paper, I have
benefitted from discussions with Sumit Das and Ashoke Sen.  The work
was supported by the SERB research grant EMR/2017/001434 of the
Government of India.

\bibliographystyle{unsrt}
\bibliography{photonprop.bib}



\end{document}